# Origin of the low precipitation hardening in magnesium alloys

C.M. Cepeda-Jiménez, M. Castillo-Rodríguez, M.T. Pérez-Prado[*]

IMDEA Materials Institute, C/ Eric Kandel, 2, 28906 Getafe, Madrid, Spain

[*]Corresponding author

**Abstract**

In this work electron backscattered diffraction (EBSD)-assisted slip trace analysis and transmission electron microscopy have been utilized to investigate the interaction of basal dislocations with precipitates in the Mg alloys Mg-1%wt.Mn-0.7%wt.Nd (MN11) and Mg-9%wt.Al-1%wt.Zn (AZ91), with the ultimate aim of determining the origin of their poor precipitation hardening. Precipitates in these alloys have a plate-shaped morphology, with plates being, respectively, perpendicular ($Mg_xNd_y$) and parallel ($Mg_{17}Al_{12}$) to the basal plane of the magnesium matrix. Mechanical tests were carried out in solid solution and peak-aged samples, in tension and compression, both at RT and at moderate temperature (250ºC). EBSD-assisted slip trace analysis revealed a clear dominance of basal slip under a wide range of testing conditions in the peak-aged MN11 and AZ91 alloys. At room temperature, the origin of the low precipitation hardening observed lies at the easiness with which precipitates are sheared by basal dislocations, due to the lattice matching at the interface with the Mg matrix. At high temperature, dislocation-precipitate interactions are highly dependent on the deformation mode. In tension, enhanced basal slip localization gives rise to high stress concentrations at the intersection between coarse slip traces and particle interfaces, leading to precipitate fracture; in compression, a more homogenous distribution of basal slip leads to the dominance of particle shearing. Our study demonstrates experimentally that basal dislocations are able to shear, and even fracture, the $Mg_xNd_y$ and $Mg_{17}Al_{12}$ plates when, for appropriate testing conditions, the local stress due to dislocation accumulation at particle interfaces exceeds the precipitate strength.

# 1. Introduction

Magnesium based alloys have received increasing interest in the last decades as a promising lightweight alternative to materials currently used in the automotive and aeronautical industries [1-5]. However, several properties of these materials, such as the corrosion resistance, the creep resistance and the formability, as well as, most notably, the room and high temperature strength, must still be improved in order to increase their commercial use [2,4,6-8].

The potential of precipitation hardening, through alloying and heat treatment, to improve the strength of Mg alloys and to reduce their mechanical asymmetry has been explored in a relatively high number of recent studies [9-13]. The results have been somewhat disappointing and it is generally agreed that the strength level that can be achieved via precipitation is still considerably lower than that obtained in counterpart aluminium alloys [4,8].

The degree of precipitate strengthening in metallic alloys is controlled by different factors such as precipitate shape, number density, and size [14], and also by the particularities of the interactions between dislocations and precipitates during deformation [4,8]. The latter are highly dependent, among other factors, on the degree of interface coherency between the particle and the matrix. Upon encountering a non-coherent precipitate, which is in general non-deformable, dislocations form Orowan loops around them; however, when there is a high degree of interface coherency with the matrix, the likelihood of a dislocation shearing the particle increases dramatically [14]. Whether or not a phase is sheared by dislocations will depend also on the degree of deformation. It is likely that incoherent or semi-coherent precipitates are not cut at small strains, but may be sheared by dislocations at larger deformations. Finally, some incoherent precipitates, such as the θ phase in Al-Cu alloys, are not cut by dislocations, but may fracture at large strains due to local stress concentrations at precipitate interfaces by dislocation accumulation [15]. Overall, the interaction mechanisms between precipitates and dislocations in magnesium alloys under different deformation conditions remain unclear.

It is worth noting that traditional modeling and theoretical calculations of the strengthening effect of precipitates interacting with basal and non-basal dislocations usually consider the hypothesis that an Orowan mechanism operates [16-19]. However, it has been recently demonstrated that certain precipitates in Mg alloys can be sheared during the deformation process [14,20-22]. For example, Wang *et al* [20] reported that MgZn rod precipitates in a Mg-5%Zn alloy were sheared by basal slip. On the other hand, Solomon *et al* [14] reported that the coherent β''' phase in a Mg-Nd alloy was sheared by basal slip dislocations and that the β' phase in a Mg-Y alloy was sheared by basal and non-basal <c+a> slip. Bhattacharyya *et al* [21] provided evidence that prismatic plate-shaped β' precipitates in the WE43 alloy can be sheared by basal dislocations, while non-basal <a> and pyramidal <c+a> dislocations are unable to shear these precipitates. In the latter case, the Orowan mechanism was reported to operate during plastic deformation, and precipitation hardening could be accurately predicted using the Orowan model. In the particular case of polycrystalline Mg alloys, it is to date not possible to accurately predict the dominant dislocation-precipitate interactions under specific testing conditions, and considerable work needs to be done in order to elucidate this matter, which is of crucial importance in order to fully exploit the potential of precipitation hardening.

Understanding dislocation-precipitate interactions requires, first, to have a reliable method to analyze the nature of the dislocations that become active under specific testing conditions in the alloy under investigation. Electron backscattered diffraction (EBSD)-assisted slip trace analysis [23,24] provides direct and statistically sound evidence of the active deformation mechanisms and has recently proven key to solving several remaining controversies regarding the plasticity of Mg alloys [25-27]. The combination of this methodology with the focused ion beam (FIB) lift-out technique for the preparation of transmission electron microscopy (TEM) thin-foil samples at selected grains with well-known orientations, provides an invaluable new tool to analyze the interaction of specific dislocations, gliding along known slip planes, with precipitates. Earlier work by the present authors on the operative deformation mechanisms in the age hardenable MN11 and AZ91 alloys, using EBSD-assisted slip trace analysis as the main characterization tool, revealed a clear dominance of basal slip under a wide range of testing conditions [28,29]. In particular, in the MN11 alloy, basal slip was reported to be the main active system at room (RT) and high temperature, in tension in both solid solution and aged samples, and in compression at high temperature in aged samples. Moreover, in the precipitate-containing MN11 microstructures, deformed in tension at high temperature, enhanced inter- and intragranular localization of basal slip were observed, leading to increased softening and, thus, triggering an inverse yield strength asymmetry [29]. In the AZ91 alloy, this method also confirmed the dominance of basal slip during deformation under tension at RT in solution treated (~82% basal slip activity) and peak-aged (~85% activity) samples [28].

The aim of the current study is investigate the interaction of basal dislocations with precipitates in the MN11 and AZ91 alloys, using EBSD-assisted slip trace analysis and TEM as the main characterization tools. Precipitates in these two materials have a plate-shaped morphology, with plates being perpendicular and parallel, respectively, to the basal plane of the magnesium matrix. These precipitates, which have been exhaustively characterized using conventional transmission electron microscopy (TEM) and selected area electron diffraction (SAED) [4,30-32], act as poor strengtheners of basal slip. This work aims to unveil the fundamental origin of such limited strengthening capability in a wide range of testing conditions.

## 2. Experimental procedure

The materials investigated in the present work are the MN11 alloy, with a chemical composition of Mg-1.0 wt% Mn-0.7 wt% Nd, and the AZ91 alloy, with a composition of Mg-9.0 wt% Al-0.70 wt% Zn. The as-received MN11 alloy was first produced by gravity casting, homogenized at 350ºC for 15h, and then machined up to a diameter of 93 mm for extrusion. Indirect extrusion was carried out at 300ºC and 5.5 mm/s to produce bars of 17 mm in diameter, which corresponds to an extrusion ratio of 1:30. The resulting microstructure, a solid solution, is termed the "as-extruded" condition. In a previous study [33] it was determined by TEM that, in the as-extruded state, Nd atoms are present mostly in solid solution. The as-extruded alloy was heat treated during 2h at 275ºC in order to induce a high level of precipitation. On the other hand, slabs of the as-received AZ91 material, approximately 8 mm in thickness, were first solution treated at 420ºC for 23 h, and then hot rolled at 420ºC using four passes, each of ~17% reduction, with an inter-pass annealing of 10 min at 420ºC. A final thickness of ~4 mm was obtained. A set of as-rolled workpieces were then solution treated at 420ºC for 18 h followed by water quenching. The resulting microstructure is termed the "solid solution" condition. In order to generate a microstructure in the "peak aged" condition, a set of solid solution-treated specimens were aged at 210ºC for 8 h. The microstructure and texture of the solid solution and peak-aged alloys under investigation have been reported earlier [28,29].

Mechanical tests were carried out in the solid solution and peak-aged MN11 samples in tension both at RT and at 250ºC, and in compression at 250ºC, with an initial strain rate of $10^{-3}$ s$^{-1}$, in order to investigate the effect of precipitation on the mechanical behavior and its interaction with dislocations. Dog-bone tensile samples of 10 mm gage length and 3×2.5 mm$^2$ transversal section were electrodischarge-machined out the as-extruded and aged bars, with the tensile axis parallel to the extrusion direction. A second set of prism-shaped compression samples, with 7 mm gage length along ED and 4×4 mm$^2$ section, were also machined out of the as-extruded and peak-aged bars.

Uniaxial tensile and compression tests were then carried out using an Instron 3384 universal electromechanical testing machine furnished with a tubular ceramic furnace. The testing temperature was monitored using a K-type thermocouple placed against the gage section of each sample. On the other hand, a series of dog-bone tensile samples with 10 mm gage length and transversal section of 2×2.5 mm$^2$ were also electrodischarge-machined out of the as-rolled AZ91 sheets and, subsequently, they were subjected to solution and peak-aging heat treatments. Tensile tests were then carried out at RT and at initial strain rates of $10^{-3}$ and $5×10^{-5}$ s$^{-1}$ using a screw-driven tensile stage (Kammrath and Weiss, Dortmund, Germany).

Two tests per sample were performed to failure under all the mentioned testing conditions with the aim of characterizing the full macro-mechanical response (yield stress (YS), maximum strength and ductility). The YS corresponding to each test was calculated as the true stress at 0.2% engineering strain. One additional test was stopped at a strain of 10% in to order to evaluate slip activity by EBSD-assisted slip trace analysis, following the experimental procedure described elsewhere [25]. Therefore, the yield strength values were averaged from three tests per selected condition. The standard deviation was found to be 8%.

In order to study the interaction between precipitates and dislocations, thin lamellae were extracted from selected grains of known orientation, where dominant dislocation types had been previously analyzed by slip trace analysis. TEM lamella extraction was carried out using a dual-beam scanning electron microscope (SEM)-FIB Helios NanoLab 600i, FEI. The SEM-FIB instrument is equipped with a field emission electron gun, and a gallium liquid metal ion source. An Omniprobe TM micromanipulator, that allows to control a tungsten needle, as well as a Pt-gas injection system, were utilized to extract TEM foils from selected locations. The site of interest is first covered with a stripe of platinum. The surrounding material is then etched away using the Ga-ion beam with a FIB acceleration voltages of 30 kV and different ranges of progressively lower ion beam currents from 65 nA to 0.79 nA, until a foil thickness of about 1.5 μm is achieved. After that the tungsten needle is used to extract and move the lamella to a V shaped copper holder, where it is fixed by Pt deposition. Then, the lamella is progressively polished from both sides by decreasing the Ga-ion beam current down to 16 pA and the voltage to 5 kV to a final thickness of about 100 nm. In order to minimize beam damage, low beam currents and low-kV settings were utilized, which resulted in increasing milling time and more homogeneous lamella thickness. The same FIB extraction methodology was carried out for preparation of all the thin foils, and thus similar beam damage has been introduced in all the samples tested under different testing conditions. Diffraction contrast imaging by bright field transmission electron microscopy (BF-TEM) and Z-contrast imaging by high angle annular dark field scanning transmission electron microscopy (HAADF) were performed using a Talos F200X FEI TEM operating at 200 kV. In addition, high-resolution transmission electron microscopy (HRTEM) images, providing an atomic level insight into the interfacial structure, and the corresponding Fast Fourier Transformations (FFT), were also obtained in order to investigate the lattice matching between precipitates and the magnesium matrix. Finally, 3D characterization of precipitate damage after tensile testing was performed by means of scanning transmission electron microscopy (STEM) computed nanotomography (nano-CT). Several videos illustrating the detailed morphology of precipitates after testing through STEM nano-CT are provided as supplementary material.

## 3. Results and discussion

*3.1. Dislocation-precipitate interactions in the MN11 alloy at room temperature.*

The true-stress-true plastic strain curves corresponding to tensile tests performed until failure at RT and at an initial strain rate of $10^{-3}$ s$^{-1}$ along the ED for the MN11 alloy in the as-extruded (solid solution) condition and after peak-aging at 275ºC for 2h are depicted in Fig. 1a. As can be clearly seen, the strengthening effect of precipitation is minimal. Indeed, both the YS and the maximum true stress remain almost invariant after age-hardening.

An earlier study [29], using EBSD-based slip trace analysis, revealed that, under tension at RT, basal slip is the dominant deformation mechanism both in the as-extruded and peak-aged conditions, and that the basal activity decreases with increasing precipitation level. The number of traces analyzed for samples tested in tension up to ~10% was 65 and 191 for the MN11 alloy in the as-extruded and peak-aged condition, respectively. The main objective of this procedure is to capture the appearance of slip traces in as many grains as possible during deformation. In the as-extruded alloy, 83% of the slip traces corresponded to basal slip, while the contribution of prismatic and pyramidal slip was 12 and 5% respectively. After age hardening the frequency of basal traces decreases to 63%, while the frequency of prismatic and pyramidal traces increases to 13 and 24%, respectively. In both cases, slip traces were observed to be homogeneously distributed throughout the gage length. As an example, Fig. 1b shows a SEM micrograph that illustrates the morphology of slip traces in the peak-aged condition.

In order to determine the origin of the poor precipitation hardening of this alloy in the peak-aged condition, the precipitate-dislocation interaction mechanisms were examined with the aid of TEM. TEM examination was also performed on the peak-aged MN11 in the undeformed state to characterize the particle distribution in the alloy without deformation and to discard any kind of damage introduced by the FIB during the lamella preparation (Fig. 2). Inspection of the undeformed sample along the [2-1-10] zone axis reveals that Nd-containing precipitates, which nucleate at the Mn particles, are plates, ~200-300 nm long, with a prismatic {1010} habit and their longer axis parallel to the matrix [0001] direction [19,33]. A number of recent studies have investigated the precipitation sequence and hardening mechanisms in Mg-Nd alloys. However, the precipitation sequence during aging of Mg-Nd alloys remains controversial [4,8,14,34].

Thus, Mg-Nd precipitates will be termed as $Mg_xNd_y$. On the other hand, despite the beam damage introduced during the lamella preparation, no apparent damage can be observed on the precipitate surface.

The experimental approach performed to characterize the precipitate-dislocation interaction mechanisms is illustrated in Fig. 3. First, a grain with visible slip traces was selected after SEM examination of the gage length surface after a strain of 10% (Fig.3a), and its orientation was measured by EBSD (Fig. 3b). Then, the assignment of an individual slip trace of the selected grain to a specific slip system was carried out by inputting its Euler angles into a MATLAB code, which provides as output a visual representation of all the possible plane traces corresponding to that particular orientation (Fig. 3c) [24]. Comparison of the slip trace under study with those simulated by the code allows selecting the actual active slip system (Fig. 3c). The main objective of this procedure is to capture the appearance of slip traces in as many grains as possible during deformation. In general only one set of parallel slip traces are detected and counted once for each grain. Therefore, using this procedure, a grain with visible basal slip traces was selected in the peak-aged MN11 alloy, tested in tension at RT and $10^{-3}$ $s^{-1}$, and a TEM foil was extracted by FIB milling from that particular grain (Fig. 3d). The TEM lamella was milled perpendicular to the basal slip traces, and nearly parallel to the prismatic (2-1-10) plane. This orientation was selected in order to analyze the interaction between basal dislocations and the plate shaped $Mg_xNd_y$ precipitates, which as mentioned present a prismatic {10-10} habit.

Figs. 4-6 contain representative TEM images from the selected grain illustrating dislocation-precipitate interactions at different magnifications. In all cases, a [2-1-10] zone axis was selected. Fig. 4a illustrates the SAED pattern corresponding to the selected grain with that particular orientation. Fig. 4b is a bright field (BF) TEM micrograph taken in that same area that constitutes clear evidence of particle shearing along the basal plane. Basal dislocations are shown to be capable of shearing a series of parallel precipitates lying at distances smaller than 50 nm from one another (see the Supplementary material (Video-Fig.4) for a clear 3D presentation of precipitate shearing with observable multiple offsets due to basal slip). The intense shearing due to the passage of multiple basal dislocations leads to an apparent misalignment of the longitudinal axis of precipitates with respect to the [0001] direction, as observed in Fig. 4b. In Fig. 4c, a HRTEM micrograph taken at higher magnification with high contrast at atomic level, it can be clearly observed that this apparent precipitate misalignment is due to multiple small steps present at the precipitate edge (red arrows) as a consequence of intense and progressive shearing by basal dislocations. Similar precipitate shearing leading to the formation of numerous small steps at the edge of another set of two precipitates is also shown in Fig. 5 (black arrows). Finally, the low magnification STEM micrograph of Fig. 6 proves that most precipitates have been sheared by gliding dislocations.

It is our contention that the origin of the low precipitation hardening observed for the MN11 alloy at RT lies at the easiness with which precipitates can be sheared by basal dislocations. This high shearability can be attributed, first, to the high degree of lattice matching between the precipitate and the matrix. Although the precipitation sequence during aging of Mg-Nd alloys has been exhaustively characterized earlier, several aspects regarding the structure of precipitates and the precipitation kinetics remain unsolved [4,8,14,34]. Recent reports [14] suggest that, upon aging Mg-Nd alloys at temperatures between 150 and 250ºC, precipitation of β''' plate-shaped precipitates with {10-10}$_{Mg}$ habit planes, such as those observed in the current work, takes place. These plates commonly form in pairs or clusters, as seen in Fig. 5. The β''' phase has an orthorhombic crystal structure, with Nd atoms lying on Mg lattice positions, and its interface is reported to be coherent with the Mg matrix. A second factor that facilitates precipitate shearing is the small cross section that must be traversed by basal dislocations. It has been reported [4] that many of the precipitates in magnesium alloys form as plates, with very small thickness, which remains stable even after long aging at high temperature. Indeed, the thickness and, for that matter, the width, of Mg$_x$Nd$_y$ precipitates, were observed to remain, respectively, at approximately 10 and 50 nm for age-hardening times as high as 20 h at 275ºC [19]. Therefore, the high degree of lattice matching with the Mg matrix, and the relatively small cross section, justify the high shearability of $Mg_xNd_y$ precipitates by basal dislocations.

*3.2. Dislocation-precipitate interactions in the MN11 alloy at high temperature.*

Fig. 7a illustrates representative tensile and compressive true stress-true plastic strain curves corresponding to peak-aged samples (275ºC for 2h) tested along ED until failure at 250ºC, in tension and compression, and at an initial strain rate of $10^{-3}$ s$^{-1}$. It can be observed that the average compressive yield stress (CYS=102 MPa) is higher than the tensile yield stress (TYS=87 MPa), resulting in a reversed YS asymmetry (CYS/TYS=1.17) [29].

As demonstrated earlier [29], basal slip is also the dominant deformation mechanism in the peak-aged MN11 alloy under the current testing conditions. The frequency of basal traces was, indeed, found similar in the sample tested in tension than in compression (75 vs. 85%, respectively). SEM examination of the appearance of basal slip traces in the peak-aged MN11 alloy tested here in tension (Fig. 7b) and in compression (Fig. 7c) shows enhanced levels of intra- and intergranular localization of basal slip in tension, in agreement with [29]. This is evidenced, first, by the presence, in the grain interiors, of fewer, but thicker, slip traces, associated with deeper surface steps (coarse slip) and, second, by basal slip concentration along bands traversing a large number of grains. Several explanations have been put forward to rationalize the origin of slip localization, a phenomenon that has been widely analyzed in Al-Li alloys [35-38]. The motion of dislocations reportedly reduces the order across the slip plane, thus decreasing the stress required to move additional dislocations. This leads to enhanced dislocation nucleation along certain slip planes (coarse slip) and to a tendency toward planar strain [35]. Extensive strain localization has been reported to give rise to local work softening and low ductility [38]. On the other hand, the easiness of dislocations to shear precipitates also leads to enhanced localization [35-37] and, therefore, in precipitate containing alloys, both mechanisms contribute to different extents, depending on the testing conditions. It seems logical that basal slip localization is promoted in Mg alloys since, due to their low symmetry, the CRSS of basal slip is significantly lower than that of other available deformation mechanisms.

In order to investigate whether the different degree of basal slip localization in tension and in compression at high temperature has any influence on dislocation-particle interactions in the MN11 alloy under study, TEM foils were extracted from grains where the active deformation mechanism was basal slip in specimens deformed under the two mentioned deformation modes up to strain of 10%. Figure 8 contains two BF TEM micrographs corresponding to the peak-aged MN11 alloy tested in tension at 250ºC and $10^{-3}$ s$^{-1}$. It can be clearly seen that, during testing, precipitates undergo localized fracture at multiple locations throughout their length. It is our contention that particle fracture in tension is caused by local stress concentrations at the intersection between coarse basal slip traces (Fig. 7b) and precipitate-matrix interfaces. This stress concentration on the precipitate interface allows to reach the critical failure stress of the precipitate at a lower macroscopic applied stress. On the other hand, no shear offset in the precipitate-matrix interface was observed after fracture as clearly viewed in the video file Video-Fig8 provided as Supplementary material. The image rotation in the nanotomography video confirms the complete fracture of precipitates under tension load and discards any kind of imaging artifact.

Figures 9 and 10 contain two BF TEM micrographs corresponding to the peak-aged MN11 alloy tested in compression at 250ºC and $10^{-3}$ s$^{-1}$. It can be clearly seen that, in compression, basal dislocations shear the precipitates and create observable offsets at precipitate-matrix interfaces. The particles in Fig.9 have been sheared multiple times along their length by basal dislocations along <11-20> directions, and multiple portions of the precipitate have been shifted with respect to each other (the large shifts observed are also perhaps due partially to the compressive hydrostatic stress). Fig. 10 provides additional sound evidence of basal dislocations shearing precipitates.

In summary, dislocation-particle interactions at high temperature are highly dependent of the deformation mode. In tension, enhanced basal slip localization gives rise to stress concentrations at particle interfaces and to their subsequent fracture; in compression, a more homogeneous distribution of basal slip leads to the dominance of particle shearing.

*3.3. Dislocation-particle interactions in the AZ91 alloy at room temperature.*

Figure 11 shows the true stress-true plastic strain response of solution treated and peak aged AZ91 specimens tested in tension until failure along the rolling direction at RT and $10^{-3}$ s$^{-1}$. The average YS corresponding to the solution treated material is 163 MPa, and this increases slightly up to ~202 MPa in the peak-aged condition. This poor strengthening response to aging has been attributed to the fact that the $Mg_{17}Al_{12}$ precipitates are not well oriented to effectively block basal slip. The predominant orientation relationship (OR) between the lattice of $Mg_{17}Al_{12}$ precipitates and the Mg matrix satisfies the Burgers OR, i.e., $(0001)_{Mg}||(011)_{Mg17Al12}$, and $[2\text{-}1\text{-}10]_{Mg}||[1\text{-}11]_{Mg17Al12}$ [22,39]. Thus, these plate-like precipitates, lying parallel to the basal plane, are reported to be very ineffective in blocking basal slip [4].

The activity of the different slip systems during tensile deformation of the AZ91 alloy along the RD was previously evaluated by EBSD-assisted slip trace analysis after a strain of 10% [28]. Basal slip was found to be the dominant deformation mechanism both in the solution treated and peak-aged conditions. The reported frequencies of basal slip traces were, respectively, 82 and 85%. These results are consistent with the weak strengthening effect of basal plates on basal dislocations.

In order to analyze the interaction between basal dislocations and the $Mg_{17}Al_{12}$ plates, a TEM lamella was extracted from a selected grain where parallel basal slip traces were observed. The TEM foil was cut parallel to the basal plane. Fig. 12 shows two BF TEM micrographs corresponding to the peak-aged AZ91 alloy tested in tension at RT and $10^{-3}$ s$^{-1}$. Fig. 12a illustrates the twisted shape of a Mg$_{17}$Al$_{12}$ plate, which appears to have been plastically sheared by the basal dislocations. Fig. 12b provides further evidence of the interaction between basal dislocations and the Mg$_{17}$Al$_{12}$ precipitates. Dislocations appear, indeed, to be arrested at precipitate interfaces. However, the observation of a clear shearing mechanism is challenging since many precipitates are thinner than the thickness of the electron transparent area of the TEM lamella, and therefore significant overlap between them is apparent. Thus, an attempt was made to enhance localization of basal slip in order to facilitate viewing of dislocation-precipitate interactions. Since it was reported earlier [27] that basal slip localization at the intra- and intergranular levels may be promoted at lower strain rates and/or higher temperatures, the peak-aged AZ91 alloy was tested in tension at RT and at an initial strain rate of $5\times10^{-5}$ s$^{-1}$. Indeed, SEM examination confirmed that tensile testing at such low strain rate led to enhanced strain localization in selected basal planes. A thin TEM foil was then milled parallel to basal planes in order to analyze the interaction between basal dislocations and precipitates. Fig. 13 shows the corresponding BF TEM micrographs, where shearing (Fig. 13a), and even fracture (Fig. 13b), of particles can be observed. These results confirm that precipitate fracture is promoted by basal slip localization.

A final attempt to better examine dislocation-particle interactions in the AZ91 alloy was carried out in the following way. A TEM lamella was extracted from a grain oriented with its basal plane parallel to the surface of the tensile coupon, in which c-axis compression during tensile deformation at RT and $5\times10^{-5}$ s$^{-1}$ led to the formation of a compression twin. A high activity of basal slip is not anticipated in the un-twinned matrix, since basal planes are oriented almost parallel to the tensile axis. However, on the contrary, extensive activation of basal slip is expected within the twin due to lattice reorientation. Fig. 14a illustrates the mentioned twin, embedded in the Mg matrix. The 56º rotation existing between the SAED patterns corresponding to the twin and the matrix confirms compression twinning along the {10-11} plane [40]. Plate precipitates, whose shape remains basically invariant outside the twin, are observed to undergo a noticeable rotation (~4º) inside the twin lath (red arrows). Similar observations were previously reported by Robson *et al* [41] and were attributed to the particles undergoing rigid body rotation due to twinning, but not to shearing [42]. Fig. 14b consists of a HRTEM image, taken at high magnification, where the presence of lines parallel to the basal planes suggest strong interaction between precipitates and basal dislocations inside the twin. Thus, Fig. 14b confirms extensive activation of basal slip inside the twin. Additionally, the $Mg_{17}Al_{12}$ plates located outside the twin exhibit more straight interfaces with the matrix (Fig. 14a) than particles inside the twin, which present wavy edges, resulting from a higher dislocation activity inside the twin and the continuous precipitate shearing by basal dislocations (Fig. 14b).

Although the $Mg_{17}Al_{12}$ plates are often described as incoherent [43], experimental evidences demonstrate a high interface coherency (lattice matching) between the plate broad surface and the matrix basal plane, or habit plane, according to the previously mentioned Burgers OR, $(0001)_{Mg}||(011)_{Mg17Al12}$, and $[2\text{-}1\text{-}10]_{Mg}||[1\text{-}11]_{Mg17Al12}$ [22,39]. In magnesium alloys, precipitates tend to form coherent interfaces, especially on the basal plane, as it is the most closely packed [44]. For instance, in Mg-Gd-Zn alloys [4] the orientation relationship between the equilibrium phase γ'' and α-Mg phases is such that $(0001)_{\gamma''}//(0001)_\alpha$ and $[10\text{-}10]_{\gamma''}//[2\text{-}1\text{-}10]_\alpha$. Thus, this phase would be fully coherent with the matrix in its habit plane, but with a relatively large misfit strain in the direction normal to the habit plane. Also, in Mg-Nd-Zn alloys, the equilibrium phase γ was reported [45] to have a face-centered cubic structure and the orientation relationship $(011)_\gamma//(0001)_\alpha$ and $[-1-11]_\gamma//[2-1-10]_\alpha$, implying a coherent matching of this phase within the habit plane.

In the current study the lattice misfit at the interface between matrix and precipitate to calculate the interface coherency degree has not been determined. However, high resolution TEM (HRTEM) imaging was used to examine the alignment level of planes at the interface between the Mg matrix and a $Mg_{17}Al_{12}$ plate, both outside (Fig. 15a) and inside (Fig. 15b) the twin depicted previously in Fig.14a. The Fast Fourier Transformation (FFT) diffractograms corresponding to the matrix or twin matrix (1), to the precipitate (2) and to both of them (3), are also shown in Fig. 15. Fig. 15a shows the perfect alignment of the matrix and precipitate FFT patterns outside the twin, which reveals an high degree of lattice matching, according to the reported Burgers OR: $(011)_\beta//(0001)_\alpha$, $[1-11]_\beta//[2-1-10]_\alpha$ [4]. On the other hand, a similarly perfect alignment between the matrix and precipitates is also observed inside the twin (Fig. 15b). It can be clearly observed that the matrix-precipitate OR inside the twin presents a 60º rotation with respect to that outside the twin. This is due to the previously mentioned 56º matrix rotation due to twinning and the additional 4º rotation of the precipitates. Therefore, inside the twin, the $(0001)_\alpha$ plane fits to another equivalent $(011)_\beta$ plane of the $[1-11]_\beta$ zone axis of precipitates in order to preserve the matching between the precipitate and matrix lattice. The observed interface lattice matching can be associated to low interface energy, which facilitates shearing. The latter is particularly facilitated in the precipitates located inside the twin, which present a smaller cross section along basal planes than the particles outside the twin, where shearing is more difficult to observe. It has indeed been reported [37] that coherent and ordered precipitates promote planar slip, concentrated on a few slip systems as a result of the local work softening on the operative slip plane, ultimately leading to shear or even fracture, depending on the degree of dislocation concentration.

Relatively conflicting evidence has been published on the capability of dislocations to shear $Mg_{17}Al_{12}$ plates in Mg alloys. Some molecular dynamics (MD) simulation studies [39] attribute the limited capability of basal dislocations to shear the equilibrium $Mg_{17}Al_{12}$ plates to the relatively large inter-precipitate spacing (about 200 nm). These authors suggest that only an elastic shear strain is produced during precipitate/dislocation interactions, irrespective of the particle size. Other recent MD work [22] reported shearing of $Mg_{17}Al_{12}$ plates by prismatic, but not by basal, dislocations, due to the comparatively much higher critical resolved shear stress (CRSS) of the former. Basal dislocations were only reported to cause local elastic straining of the precipitates. Our study demonstrates experimentally that basal dislocations are able to shear, and even fracture, the $Mg_{17}Al_{12}$ plates when, for appropriate testing conditions, the local stress due to dislocation accumulation at particle interfaces exceeds the precipitate YS.

3.4. *Design of high strength magnesium alloys via precipitation hardening*

EBSD-assisted slip trace analysis of aged MN11 and AZ91 samples proved that basal slip is the main deformation mechanism under a wide range of testing conditions. Therefore, from the perspective of precipitation strengthening, mainly the interaction mechanism between precipitates and basal dislocations should be taken into consideration when designing new Mg alloys.

The current experimental observations have demonstrated that basal dislocations may be able to shear the precipitates since, due to their excellent crystallographic matching with the Mg matrix, they are not required to change glide planes at the interphase boundary. In addition, once the first dislocations shear a given precipitate, the effective particle cross sectional area becomes weaker, and this facilitates cutting of the particle on the same slip plane by incoming dislocations. Depending of the degree of basal slip localization, which is a function of testing conditions such as temperature and strain rate, precipitates may be even fractured by coarse basal slip. The latter may, additionally, promote local stress concentrations at grain boundaries, leading to premature transgranular or intergranular failure [46].

Therefore, developing high strength Mg alloys would, firstly, require precipitation of nonshearable particles containing incoherent interfaces, in order to trigger Orowan looping of basal dislocations. The appropriate approach to introduce strong incoherent precipitates into the magnesium matrix is still to be defined. Modification of the precipitate-matrix interface energy by the addition of trace elements [47] could constitute a viable alternative to significantly improve mechanical behavior. Secondly, in order to optimize strength, precipitates must, additionally, be finely dispersed in the matrix, and this may be facilitated by providing artificial nucleation sites such as dislocations and/or vacancy clusters.

Finally, basal slip localization must be prevented by introducing non-shearable particles or solutes that are able to provide impenetrable barriers to dislocation movement and, thus, contribute to break up and disperse coplanar slip, thereby ensuring a homogenous distribution of stresses during deformation [36].

In summary, the present study suggests that designing effective precipitation strengthening strategies for advanced Mg alloys will require an in-depth knowledge of the interaction between dislocations and precipitates as well as of the intra- and intergranular distribution of basal slip in order to control the magnitude of local stresses at precipitate interfaces and, thus, their deformation mechanisms.

## 5. Conclusions

In this work the dislocation-precipitate interactions in the peak-aged MN11 and AZ91 alloys have been analyzed by the combination of the EBSD-assisted slip trace analysis with transmission electron microscopy (TEM), with the aim of investigating the origin of the poor precipitation hardening observed in these alloys. With that purpose, thin lamellae were extracted from selected grains of known orientation, where dominant dislocation types had been previously analyzed by slip trace analysis. The following conclusions can be drawn from the present study:

1. In the MN11 alloy, deformed under tension at RT, basal slip is the dominant deformation mechanism both in the as-extruded and peak-aged conditions. Transmission electron microscopy (TEM) provides unequivocal evidence that the plate shaped $Mg_xNd_y$ precipitates are sheared by basal dislocations, which explains the low precipitation hardening observed for the MN11 alloy at RT.

2. At high temperature, basal slip is also the dominant deformation mechanism in the peak-aged MN11 alloy during tension and compression straining. In tension, enhanced basal slip localization gives rise to high stress concentrations at the intersection between coarse slip traces and particle interfaces, leading to precipitate fracture; in compression, a more homogenous distribution of basal slip leads to the dominance of particle shearing. The extensive strain localization in tension and the precipitate fracture is the origin of the high temperature reversed yield asymmetry in the MN11 alloy.

3. In the AZ91 alloy, basal slip was, again, found to be the dominant deformation mechanism both in the solution treated and peak-aged conditions under tension at RT. TEM examination confirms shearing and even fracture of the $Mg_{17}Al_{12}$ plates when basal slip localization is promoted at lower strain rate.

4. The high shearability of precipitates by basal dislocations is attributed to the high degree of lattice matching with the matrix. The transition from shear to fracture is shown to depend on the degree of slip localization which, in turn, is influenced by the testing conditions (temperature, strain rate, deformation mode).

5. The combined use of EBSD-assisted slip trace analysis and analytical TEM provides an invaluable tool to analyze the interaction of dislocations gliding along known slip planes with precipitates in polycrystalline alloys.

**Acknowledgments**

The research leading to these results has received funding from Madrid region under programme S2013/MIT-2775, DIMMAT project.

**Figure Captions**

Figure 1. a) Tensile true stress-true plastic strain curves corresponding to the as-extruded and peak-aged MN11 alloy tested along the ED at RT and at an initial strain rate of $10^{-3}$ s$^{-1}$. b) SEM micrograph illustrating the diffuse appearance of basal slip traces in the peak-aged MN11 alloy after tensile testing along ED at RT.

Figure 2. BF-TEM micrograph showing the orientation and morphology of Mg$_x$Nd$_y$ precipitates in the peak-aged MN11 alloys in the undeformed state. The SAED pattern corresponding to the [2-1-10] zone axis is shown in the inset.

Figure 3. (a,b,c) Methodology used for slip trace analysis and (d) TEM lamella sectioning. a) SEM micrograph corresponding to the peak-aged MN11 alloy tensile tested at RT and $10^{-3}$ s$^{-1}$, illustrating the trace under analysis in the selected grain; b) Post-test EBSD inverse pole figure map in the normal direction, from which the orientation of the selected grain is measured; c) Calculation of the 12 possible traces for the selected grain, and assignment of the trace to basal slip; d) Single HCP cell showing the orientation of the selected grain and that of the sectioned lamella, which was milled perpendicular to the basal plane and parallel to the plate precipitates in this Mg alloy.

Figure 4. Dislocation-particle interactions in the peak aged MN11 alloy deformed in tension at RT and $10^{-3}$s$^{-1}$. a) SAED pattern corresponding to the [2-1-10] zone axis; b) BF-TEM and c) HRTEM micrographs illustrating shearing of the Mg$_x$Nd$_y$ precipitates by basal slip. The white line in (c) highlights the orientation of a basal plane and the small red arrows indicate multiple shearing offsets at the precipitate edge. See video as a supplementary material.

Figure 5. Dislocation-particle interactions in the peak aged MN11 alloy deformed in tension at RT and $10^{-3}$s$^{-1}$, viewed along the [2-1-10] zone axis (Fig 3a). BF-TEM micrograph illustrating a high concentration of basal dislocations close to the Mg$_x$Nd$_y$

plates, as well as precipitate shearing. Black arrows highlight the position of basal dislocations.

Figure 6. Dislocation-particle interactions in the peak aged MN11 alloy deformed in tension at RT and $10^{-3}s^{-1}$, viewed along the [2-1-10] zone axis (Fig. 3a). Low magnification STEM micrograph showing that most $Mg_xNd_y$ precipitates were sheared by basal slip.

Figure 7. a) Tensile and compression true stress-true plastic strain curves corresponding to the peak-aged MN11 alloy tested along the ED at 250ºC and at an initial strain rate of $10^{-3}$ $s^{-1}$. b) and c) SEM micrographs illustrating the appearance of basal slip traces in the peak-aged MN11 after b) tensile and c) compressive testing at 250ºC.

Figure 8. Dislocation-particle interactions in the peak aged MN11 alloy deformed in tension at 250ºC and $10^{-3}s^{-1}$, viewed along the [2-1-10] zone axis. BF-TEM micrographs illustrating $Mg_xNd_y$ precipitates that were fractured by basal coarse slip (black arrows indicate fractures). The black lines highlight the orientation of basal planes. See video as a supplementary material.

Figure 9. Dislocation-particle interactions in the peak aged MN11 alloy deformed in compression at 250ºC and $10^{-3}s^{-1}$, viewed along the [2-1-10] zone axis. BF-TEM micrographs illustrating $Mg_xNd_y$ precipitates that were sheared by basal slip.

Figure 10. Dislocation-particle interactions in the peak aged MN11 alloy deformed in compression at 250ºC and $10^{-3}s^{-1}$, viewed along the [2-1-10] zone axis. BF-TEM micrographs illustrating the interaction between basal dislocations and $Mg_xNd_y$ plates, which lead to shearing of the precipitates. Black lines highlight the position of basal dislocations crossing precipitates.

Figure 11. Tensile true stress-true plastic strain curves corresponding to the solid solution and peak-aged AZ91 alloy tested along the RD at RT and at an initial strain rate of $10^{-3}$ $s^{-1}$.

Figure 12. Dislocation-particle interactions in the peak aged AZ91 alloy tested in tension at RT and $10^{-3} s^{-1}$, viewed along the [0001] zone axis. BF-TEM micrographs illustrate plate-shaped morphology of $Mg_{17}Al_{12}$ precipitates, being parallel to the basal plane of the magnesium matrix. The red arrows highlight basal dislocation lines reaching the particle interfaces.

Figure 13. Dislocation-particle interactions in the peak aged AZ91 alloy tested in tension at RT and $5\times10^{-5}$ $s^{-1}$, viewed along the [1-21-3] zone axis. BF-TEM micrographs illustrating (a) shearing and (b) fracture of $Mg_{17}Al_{12}$ precipitates by basal slip.

Figure 14. Dislocation-particle interactions in the peak aged AZ91 alloy tested in tension at RT and $5\times10^{-5}$ $s^{-1}$, viewed along the [11-20] zone axis. (a) BF-TEM image of a compression twin in a basal grain of the peak-aged AZ91 alloy. The insets show the SAED patterns corresponding to the matrix and the twin. (b) HRTEM image showing the intense interaction between basal slip and the $Mg_{17}Al_{12}$ precipitates inside the twin and multiple shearing offsets at the precipitate edge.

Figure 15. HRTEM images of interfaces between the Mg matrix and a $Mg_{17}Al_{12}$ precipitate in the peak aged AZ91 alloy tested in tension at RT and $5\times10^{-5}$ $s^{-1}$ and FFT diffractograms corresponding to (1) the matrix ($[2-1-10]_{Mg}$ zone axis), (2) a $Mg_{17}Al_{12}$ plate ([1-11] zone axis) and (3) both, matrix and precipitate, a) out and b) inside the twin in Fig. 14a.

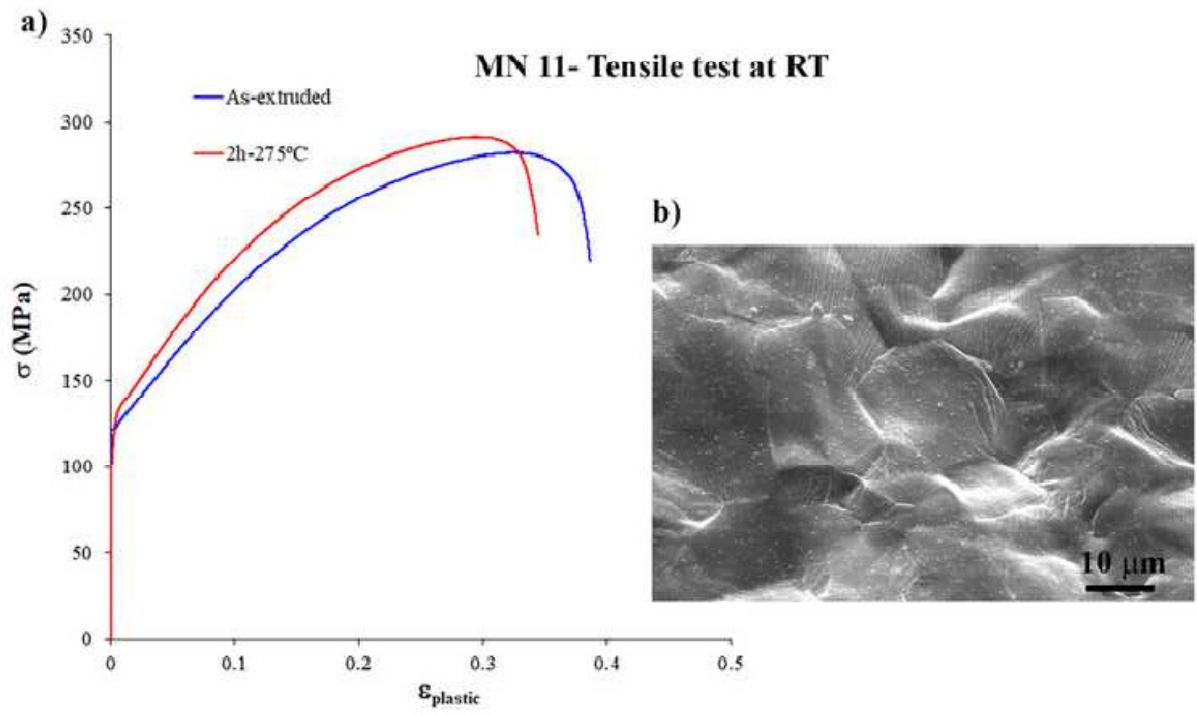

Figure 1

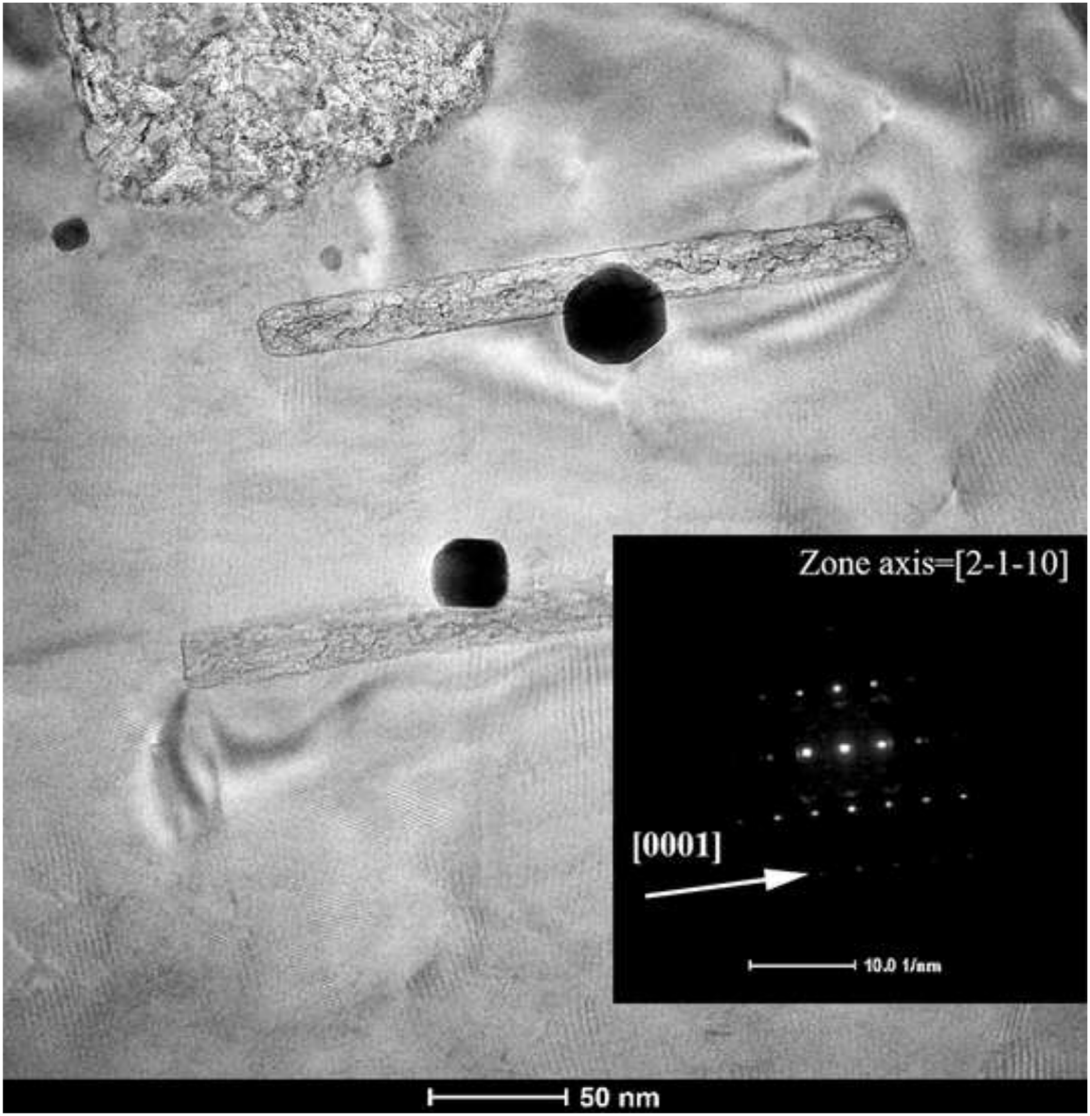

Figure 2

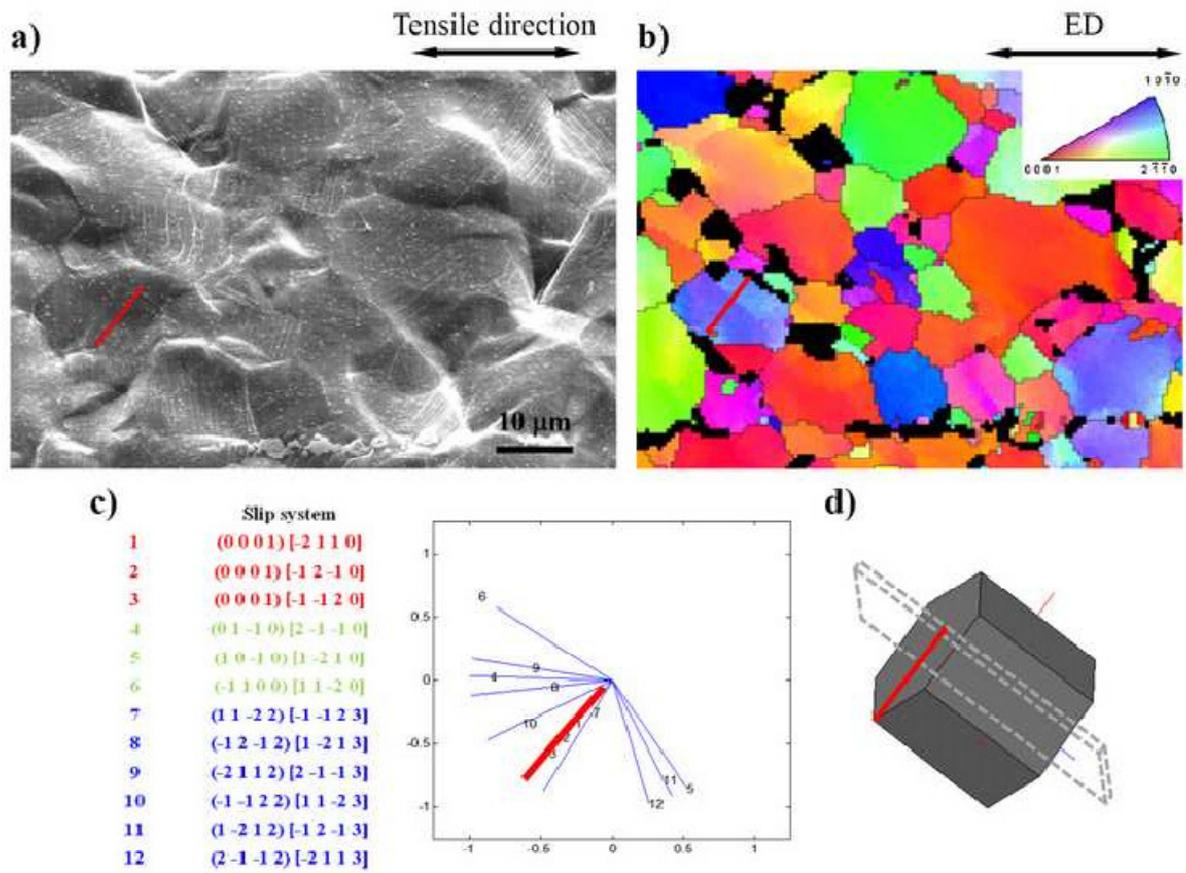

Figure 3

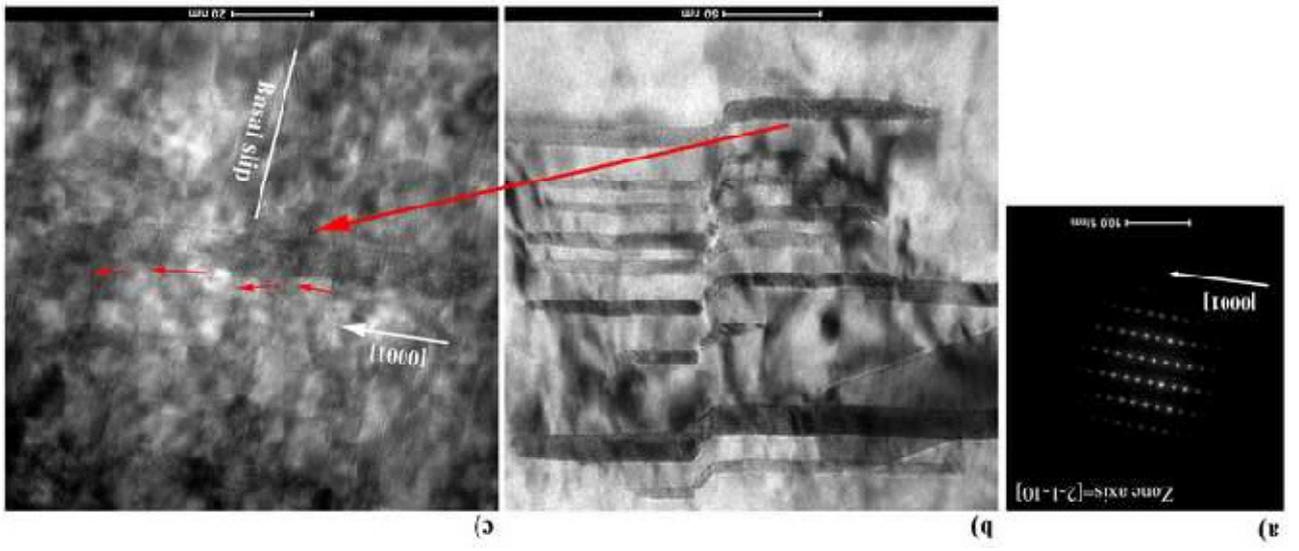

Figure 4

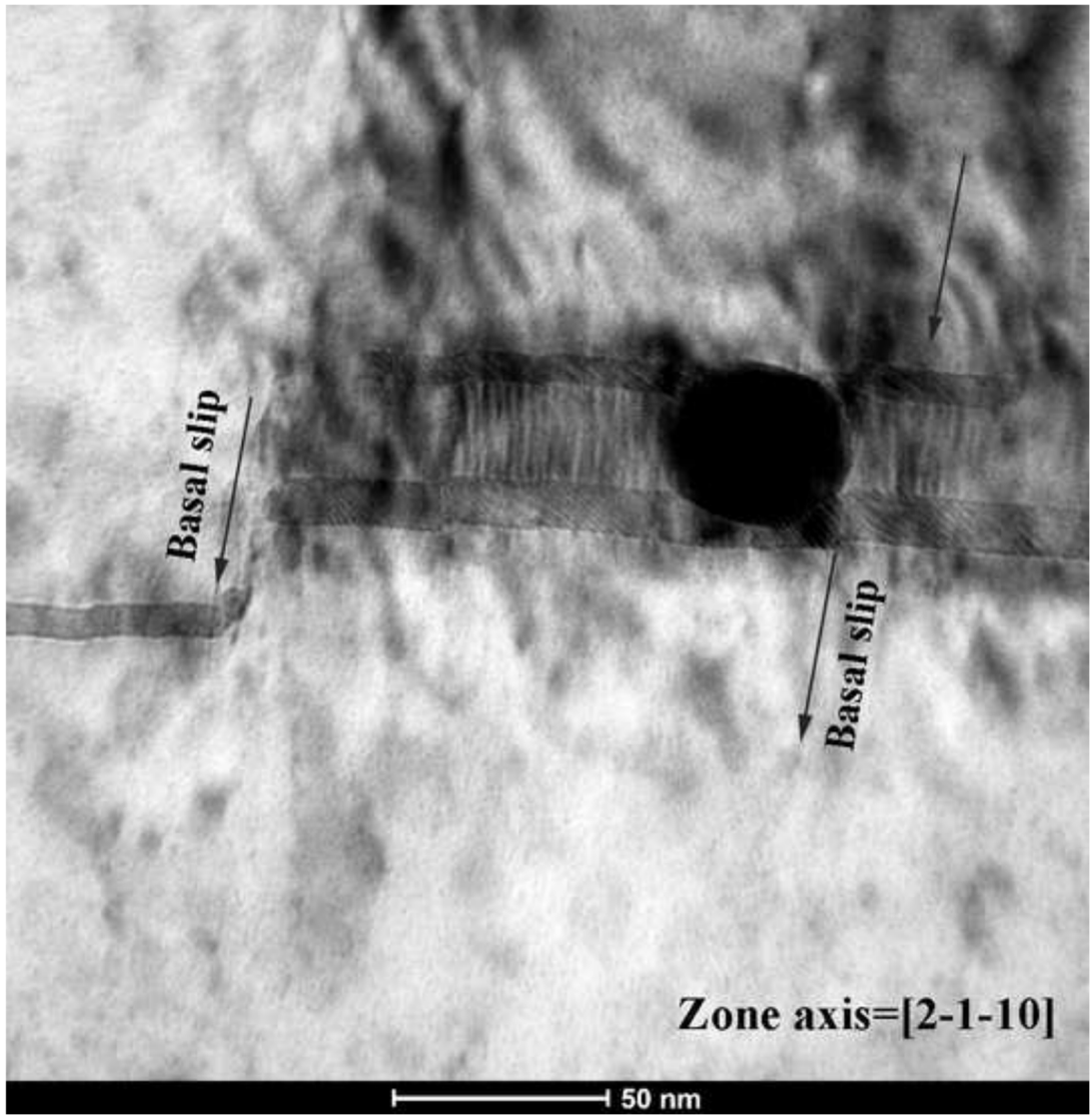

Figure 5

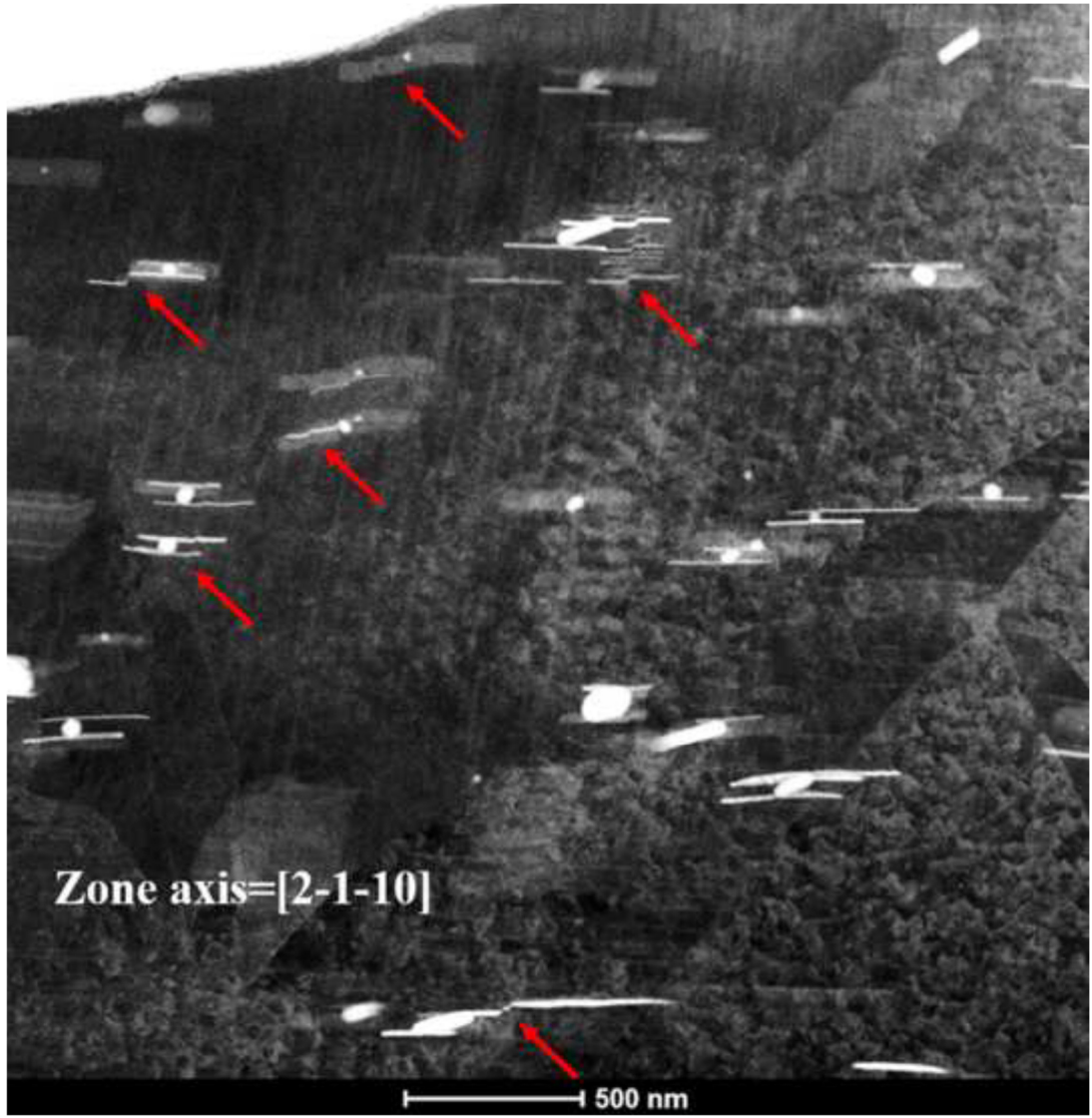

Figure 6

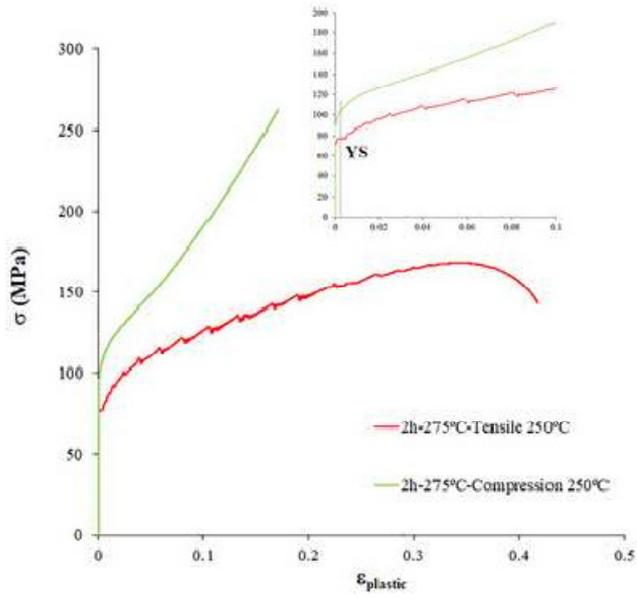
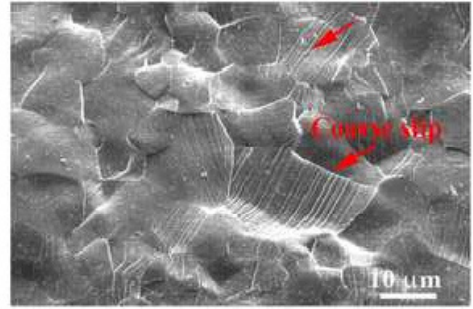
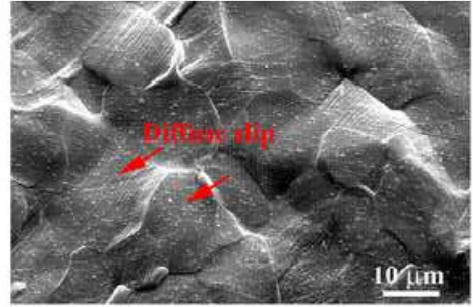

Figure 7

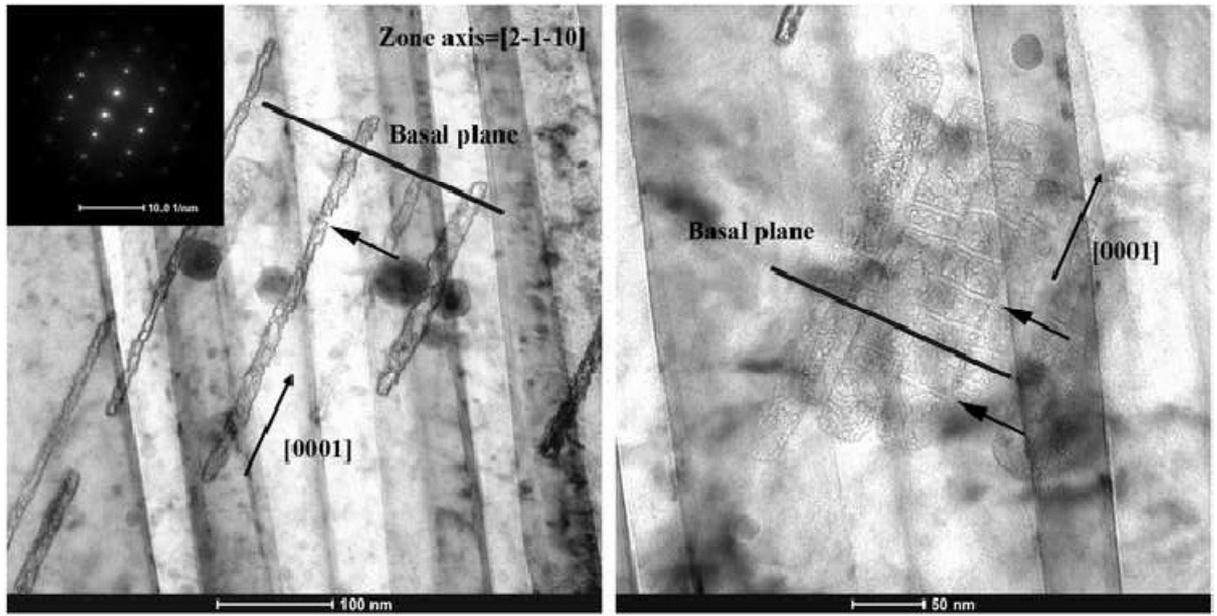

Figure 8

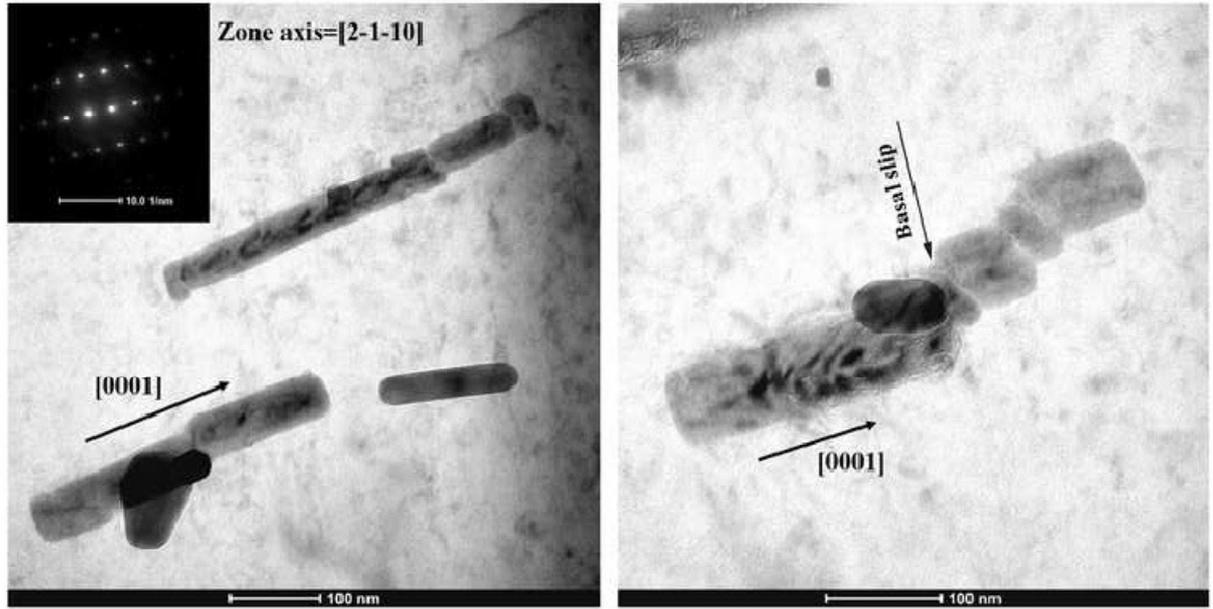

Figure 9

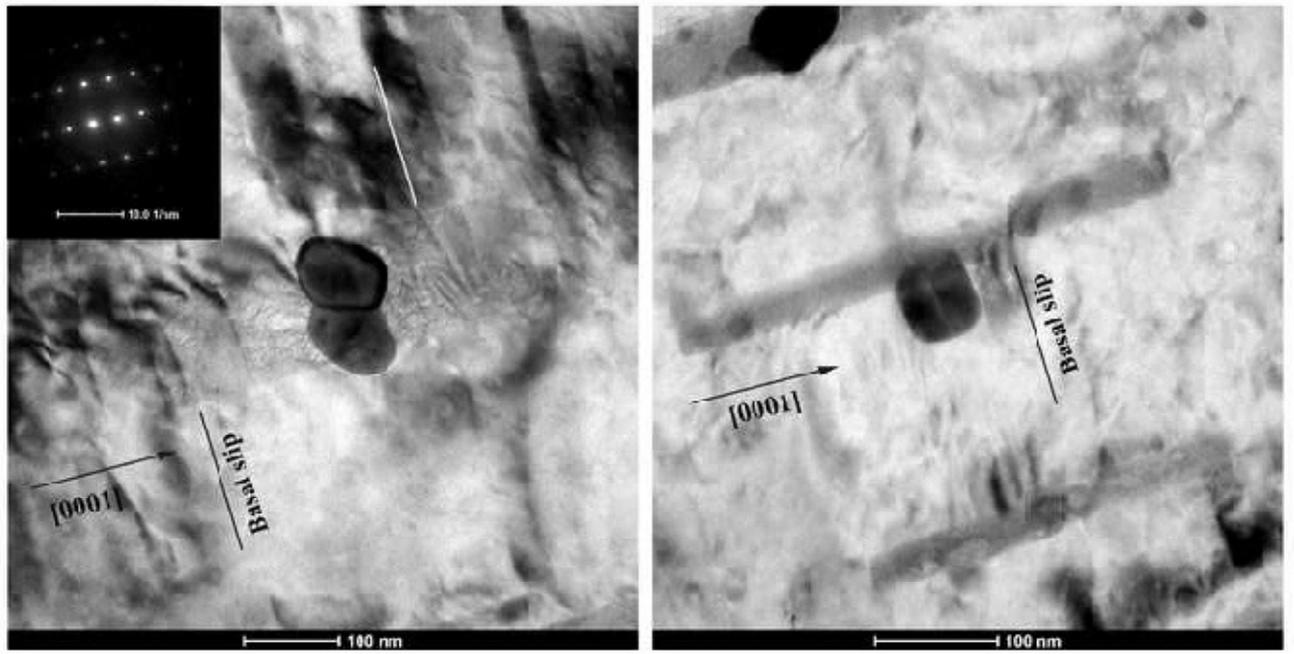

Figure 10

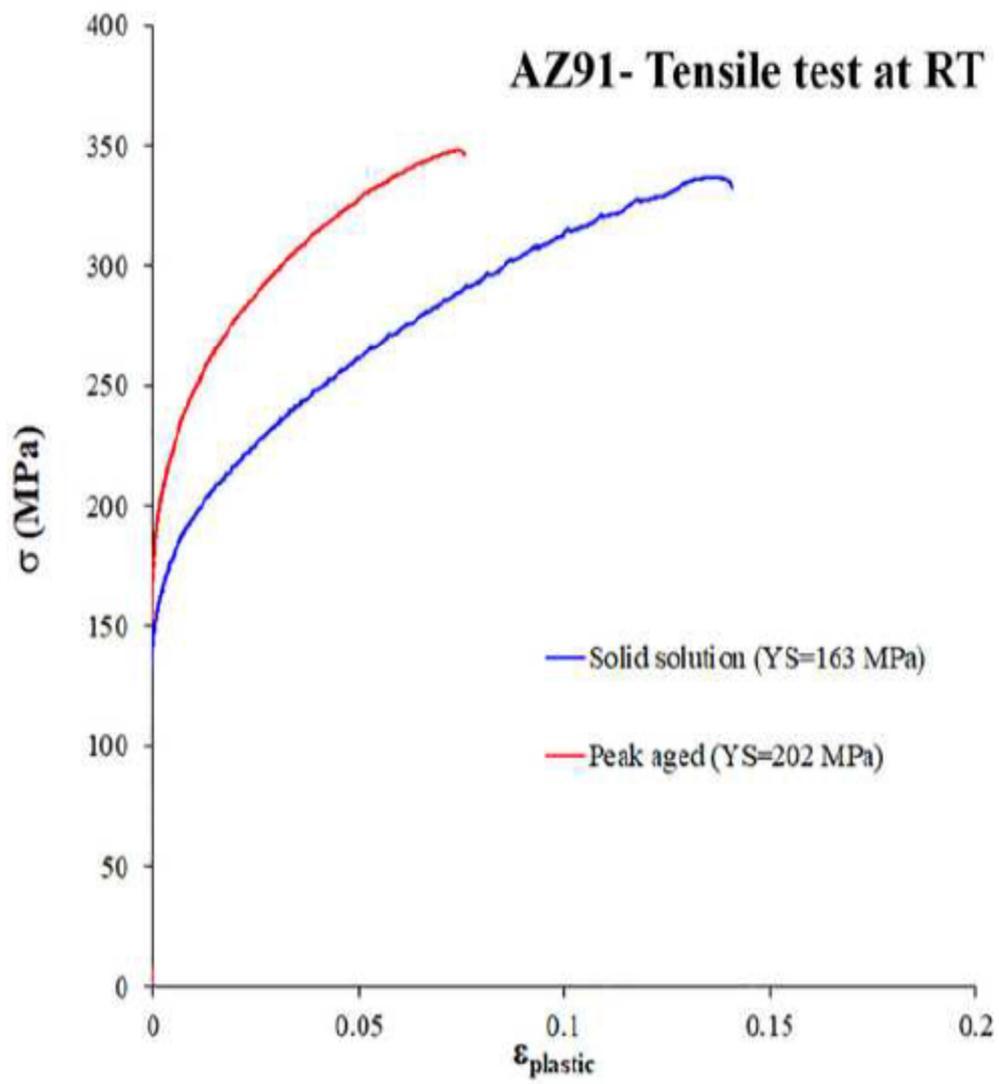

Figure 11

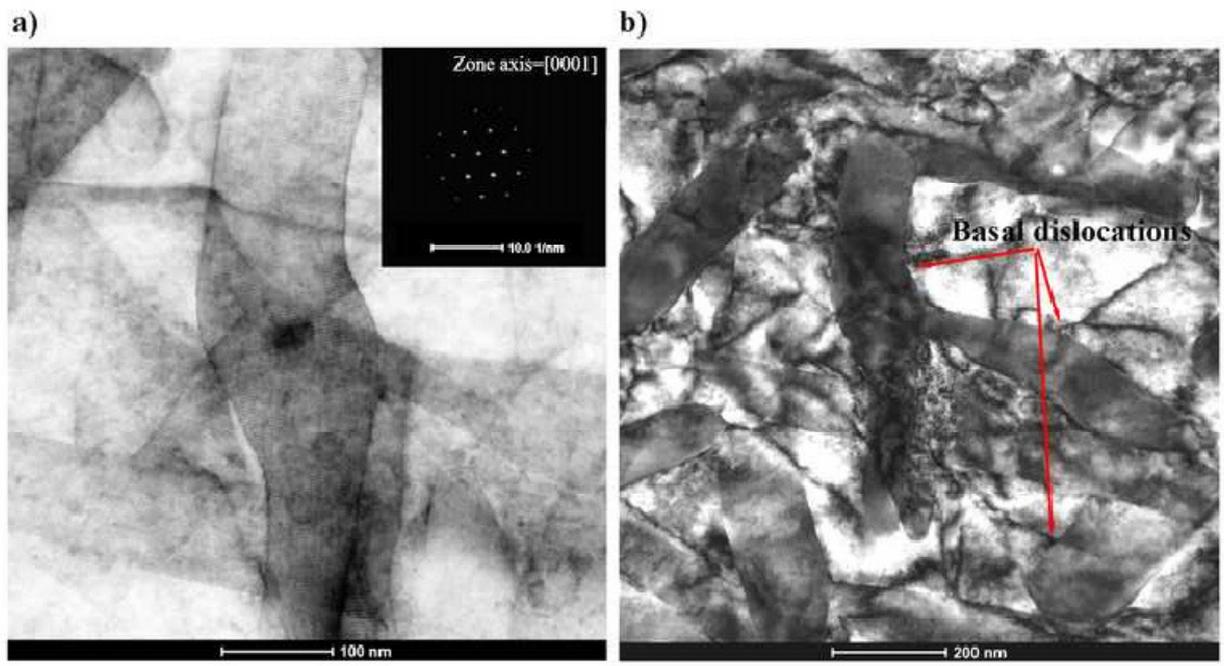

Figure 12

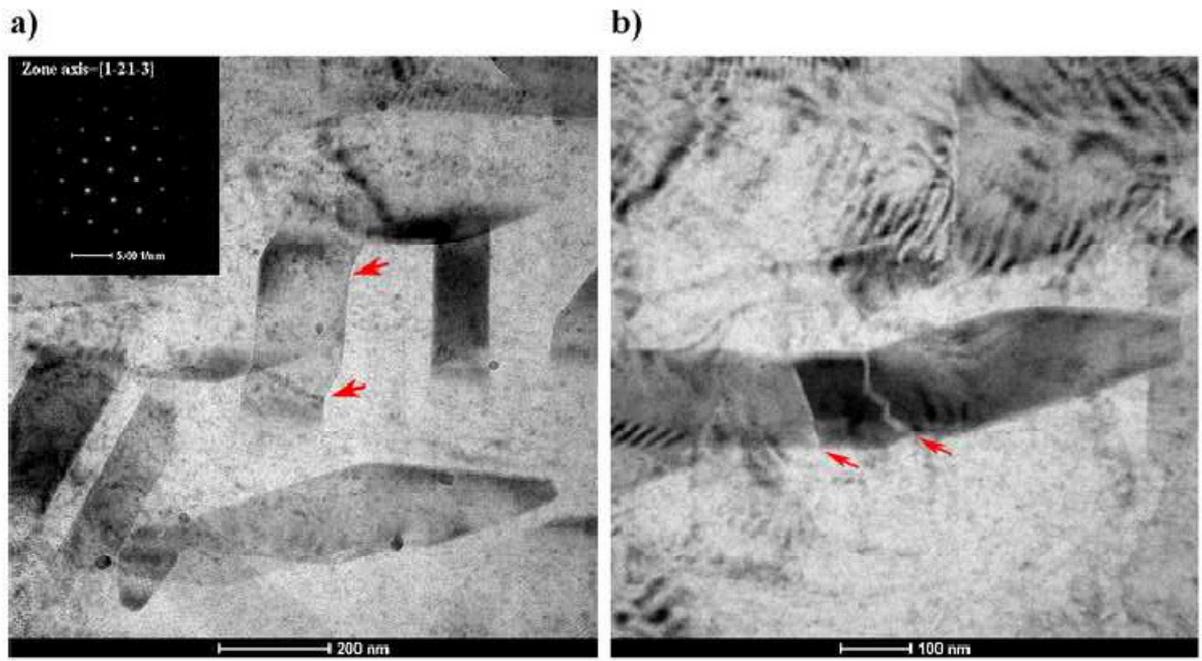

Figure 13

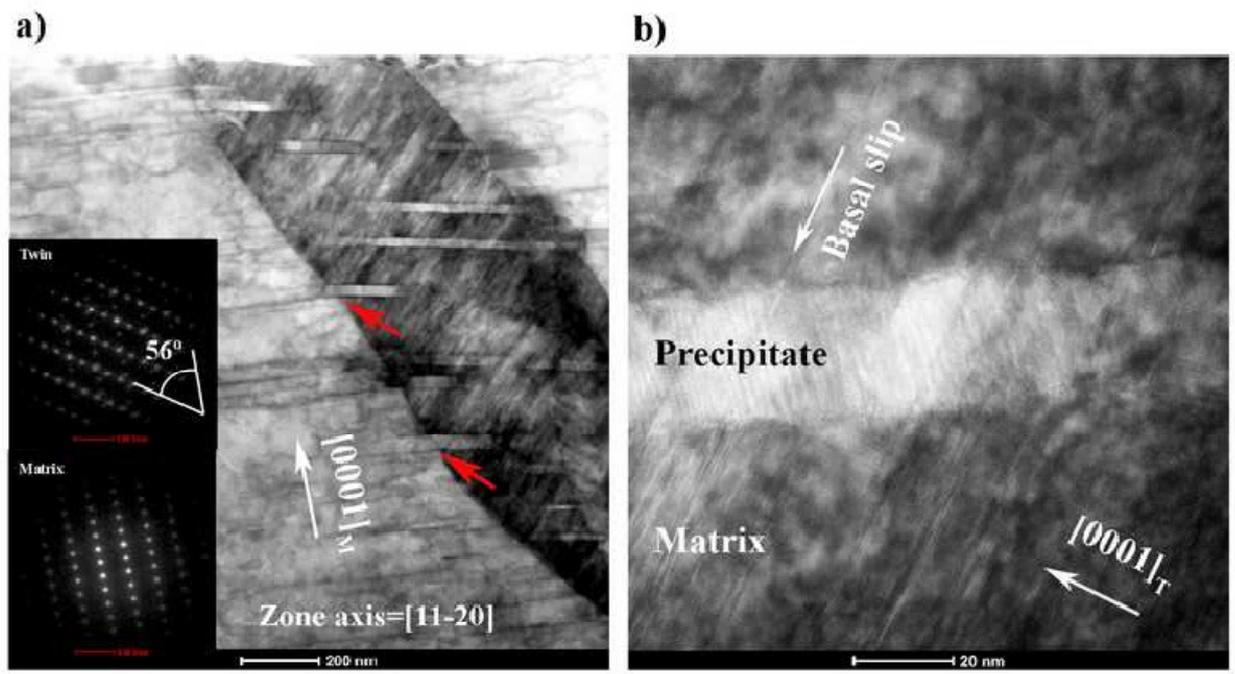

Figure 14

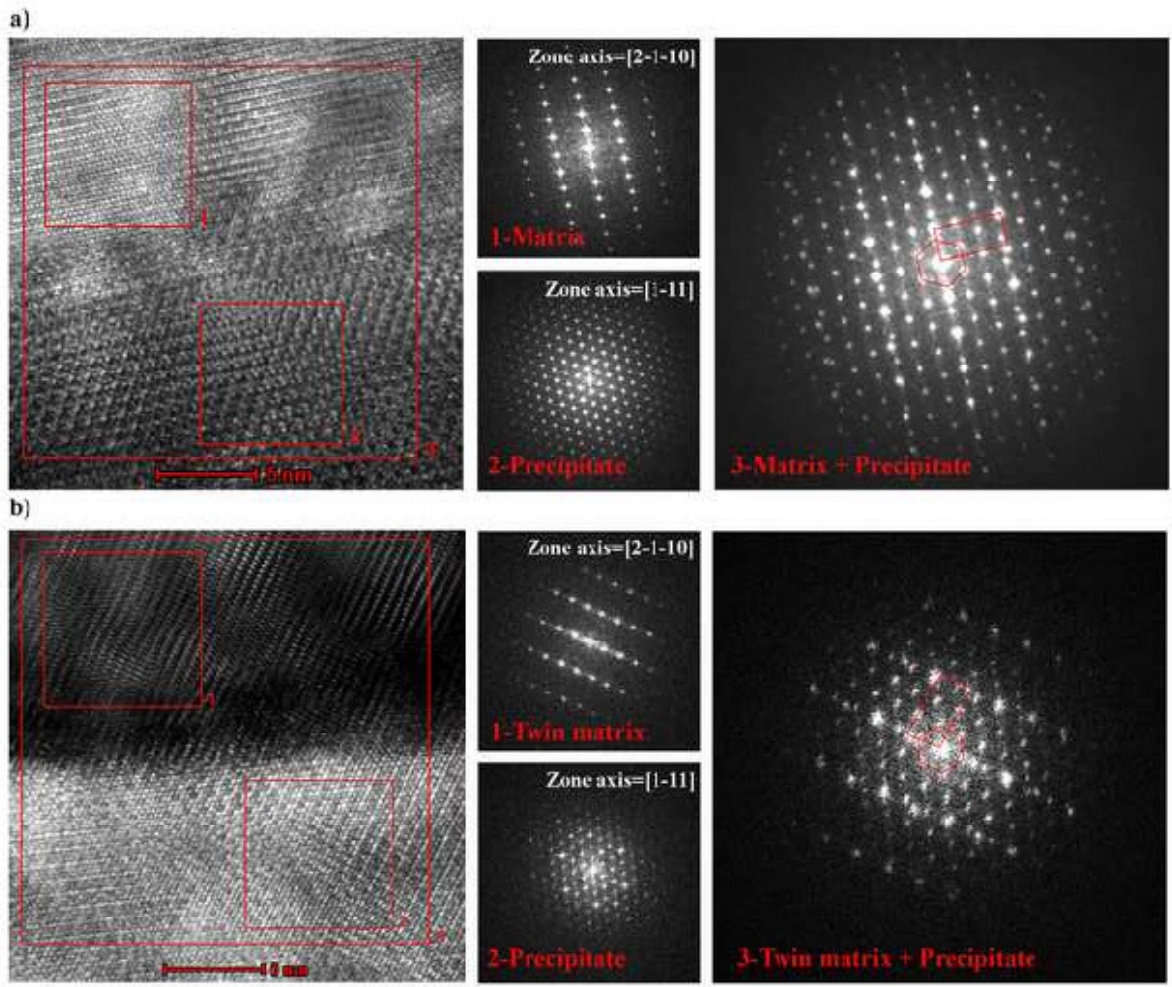

Figure 15